\documentclass[aps,prl,preprintnumbers,amsmath,amssymb,twocolumn,floatfix,superscriptaddress,nofootinbib]{revtex4-1}

\usepackage{multirow}
\usepackage{graphicx}
\usepackage{amsmath}
\usepackage{amsfonts}
\usepackage{amssymb}
\usepackage{slashed} 
\usepackage{color}
\usepackage{hyperref}
\usepackage{units}
\usepackage{braket}
\usepackage{siunitx}

\definecolor{oucrimsonred}{rgb}{0.6, 0.0, 0.0}
\definecolor{persianblue}{rgb}{0.11, 0.22, 0.73}
\definecolor{forestgreen}{rgb}{0.13,0.35,0.13}
 \hypersetup{colorlinks, citecolor=oucrimsonred, linkcolor=persianblue, urlcolor=oucrimsonred}
\newcommand{\be}{\begin{equation}}
\newcommand{\ee}{\end{equation}}
\newcommand{\bea}{\begin{eqnarray}}
\newcommand{\eea}{\end{eqnarray}}

\begin{document}

\title{Neutron Star Heating in Dark Matter Models for Muon g-2 with Scalar Lepton Partners up to the TeV Scale
}

\newcommand{\SISSA}{\affiliation{Scuola Internazionale Superiore di Studi Avanzati (SISSA), via Bonomea 265, 34136 Trieste, Italy}}
\newcommand{\INFNTrieste}{\affiliation{INFN, Sezione di Trieste, via Valerio 2, 34127 Trieste, Italy}}
\newcommand{\IFPU}{\affiliation{Institute for Fundamental Physics of the Universe (IFPU), via Beirut 2, 34151 Trieste, Italy}}
\newcommand{\NIP}{\affiliation{National Institute of Physics, University of the Philippines Diliman, Quezon City 1101, Philippines}}
\newcommand{\NTHU}{\affiliation{Department of Physics, National Tsing Hua University, Hsinchu, 30013, Taiwan}}

\author{Jan Tristram Acu\~{n}a}
\email[Electronic address: ]{jtacuna@gapp.nthu.edu.tw}

\SISSA
\INFNTrieste
\IFPU
\NIP
\NTHU

\author{Patrick Stengel}
\email[Electronic address: ]{pstengel@fe.infn.it}

\SISSA
\INFNTrieste
\IFPU

\author{Piero Ullio}
\email[Electronic address: ]{ullio@sissa.it}

\SISSA
\INFNTrieste
\IFPU

\date{\today}
\begin{abstract}
We investigate the kinetic heating of neutron stars due to dark matter scattering in minimal models with scalar lepton partners mediating interactions which can account for the anomalous magnetic moment of the muon. In particular, such models in which the dark matter is a thermally produced Majorana fermion can be extremely difficult to detect at conventional searches. We demonstrate that future infrared observations of an old neutron star population could definitively probe the entire parameter space of this scenario.
\end{abstract}
\maketitle
The hint for new physics indicated by recent measurements of the anomalous magnetic dipole moment of muon, $g_\mu -2$,~\cite{Muong-2:2021ojo} provides a starting point for the consideration of beyond-the-Standard Model (BSM) frameworks potentially accounting for the $g_\mu -2$  and embedding a dark matter (DM) candidate. Some minimal scenarios address these two aspects in terms of the very same BSM states; even in such setups, however, it has been difficult to single out other observables to confirm or reject the model. In this Letter, we consider the effect of kinetic heating of neutron stars (NSs) by DM particles in the galactic halo~\cite{Baryakhtar:2017dbj,Raj:2017wrv,Bell:2018pkk,Acevedo:2019agu}, and propose it as a generic tool to test minimal BSM frameworks in which the DM state, a thermal relic from the early Universe, is invoked, together with another state carrying electric and muonic lepton charges, to provide the required extra 1-loop contribution to the $g_\mu -2$.

To prove this connection, as a reference minimal scenario, we will consider a model in which the DM is a SM singlet Majorana fermion with DM-SM interactions mediated by chirally mixed charged scalar lepton partners. While this choice selects a small subset of the particle content of the Minimal Supersymmetric Standard Model (MSSM), respectively the bino and two sleptons (MSSM jargon will be borrowed hereafter), most MSSM realizations rely on several extra ingredients, with different DM and $g_\mu -2$ phenomenologies, and a feeble link  between the two. The model at hand (see Ref.~\cite{Ghosh:2022zef} for a recent overview) displays regions in parameter space representative of the two possible effects driving the relic density for leptophilic DM models: an enhanced bino annihilation rate when the left-right slepton mixing is sizable, as for the so-called Incredible Bulk scenario~\cite{Fukushima:2014yia}, and large co-annihilations for a relatively small mass splittings between the bino and lightest slepton (e.g. see Ref.~\cite{Acuna:2021rbg}, hereafter referred to as ASU). 

Leptophilic models have DM-nucleon interactions generated at 1-loop level, reducing the prospects for direct detection. Most relevant for NS heating are tree-level DM-lepton scatterings, since, although neutrons are the primary component of NSs, the transfer of kinetic energy can also occur by DM scattering off subdominant proton, electron and muon constituents~\cite{Bell:2019pyc,Garani:2019fpa,Joglekar:2019vzy,Joglekar:2020liw}. Moreover, due to the gravitational acceleration of DM to semi-relativistic speeds prior to scattering off NS constituents, interactions which are typically suppressed by the velocity or momentum of the DM in terrestrial direct detection experiments can potentially be relevant for NSs. 

The DM-induced NS temperatures are potentially detectable by infrared telescopes such as the James Webb Space Telescope (JWST)~\cite{Gardner:2006ky}. Recent work in Ref.~\cite{Hamaguchi:2022wpz} has investigated NS heating in similar DM models which satisfy $g_\mu -2$, including benchmark models with slepton-mediated bino interactions. Our work studies a more comprehensive parameter space encompassing such models, thus allowing more general conclusions: We demonstrate that, for both mechanisms responsible for the bino relic abundance, the scenario could be confirmed with future observations of heated NSs by telescopes such as the JWST, or alternatively, the lack of such observations could definitively exclude the full broad class of slepton mediator models.

\medskip
{\it Dark matter model.}---We define the interactions between  the bino $\Tilde{B}^0$, SM leptons and the associated scalar lepton partners $\Tilde{\ell}$,
\begin{eqnarray}
    \mathcal{L}_{\rm int} &=& - \lambda_{\Tilde{\mu}_R} e^{- \imath \phi_{\Tilde{\mu}}/2} \Tilde{\mu}_R^* \Bar{\Tilde{B}}^0 P_R \mu
    - \lambda_{\Tilde{\mu}_L} e^{ \imath \phi_{\Tilde{\mu}}/2} \Tilde{\mu}_L^* \Bar{\Tilde{B}}^0 P_L \mu \nonumber \\
    && - \lambda_{\Tilde{\nu}} e^{ \imath \phi_{\Tilde{\mu}}/2} \Tilde{\nu}_\mu^* \Bar{\Tilde{B}}^0 P_L \nu_\mu + {\rm h.c.}
    \label{eq:Lint}
\end{eqnarray}
where $P_{R,L}$ are the right- and left-handed projectors. We have introduced the $SU(2)_L$ singlet scalar muon partner $\Tilde{\mu}_R$ and the corresponding $SU(2)_L$ doublet $\Tilde{\ell}_L = (\Tilde{\nu}_\mu, \Tilde{\mu}_L)^T$. All BSM states are charged under a $\mathbb{Z}_2$ symmetry, which stabilizes the lightest such state, assumed to be the bino. For simplicity, although in principle couplings can be independent, we assume MSSM-like relations, with $\lambda_{\Tilde{\nu}} = \lambda_{\Tilde{\mu}_L}$ and $\lambda_{\Tilde{\mu}_{R,L}}  = \bar{\lambda} \sqrt{2} g' Y_{R,L}$, where $Y_{R,L}$ are the SM hypercharge
and $g'$ is the hypercharge coupling, and the extra overall factor $\bar{\lambda}$ would be equal to 1 in the MSSM. We also allow for the coupling constants to have a relative $CP$-violating phase $\phi_{\Tilde{\mu}}$. 

In order for the model described above to satisfy $g_\mu -2$, the left- and right-handed chiral smuon eigenstates must mix. The smuon mass eigenstates are related to the chiral eigenstates by the left-right mixing angle $\theta_{\Tilde{\mu}}$,
\begin{equation}
\begin{pmatrix}
\Tilde{\mu}_1  \\
\Tilde{\mu}_2 
\end{pmatrix}
=
\begin{pmatrix}
\cos \theta_{\Tilde{\mu}} & -\sin \theta_{\Tilde{\mu}}\\
\sin \theta_{\Tilde{\mu}} & \cos \theta_{\Tilde{\mu}}
\end{pmatrix}
\begin{pmatrix}
\Tilde{\mu}_L  \\
\Tilde{\mu}_R 
\end{pmatrix} \, ,
\end{equation}
where we adopt the convention in which $\Tilde{\mu}_1$ is always lighter than $\Tilde{\mu}_2$ and the left-right mixing angle is defined on the interval $[- \pi/2, \pi/2 )$. Also, to characterize the off-diagonal term in the corresponding smuon mass matrix, it is convenient to define the smuon mass-splitting parameter 
\begin{equation}
    y \equiv \left( m_{\Tilde{\mu}_2}^2 - m_{\Tilde{\mu}_1}^2 \right) \sin \left(2 \theta_{\Tilde{\mu}}\right) / \left(4 m_W^2 \right) \, ,
\end{equation}
where $m_W$ is the \textit{W}-boson mass. In models which assume minimal flavor violation such as typical realizations of the MSSM, $y \ll 1$ since the off-diagonal term in smuon mass (squared) matrix is $\propto m_\mu^2 / m_W^2$. We take the sneutrino mass to be defined as in the the MSSM,
\begin{equation}
    m_{\Tilde{\nu}_\mu}^2 \equiv m_{\Tilde{\mu}_1}^2 - m_W^2 + \left( m_{\Tilde{\mu}_2}^2 - m_{\Tilde{\mu}_1}^2 \right) \left[ 1- \cos \left(2 \theta_{\Tilde{\mu}}\right) \right] / 2 \, .
\end{equation}

\begin{figure}[t]
\includegraphics[width=\linewidth]{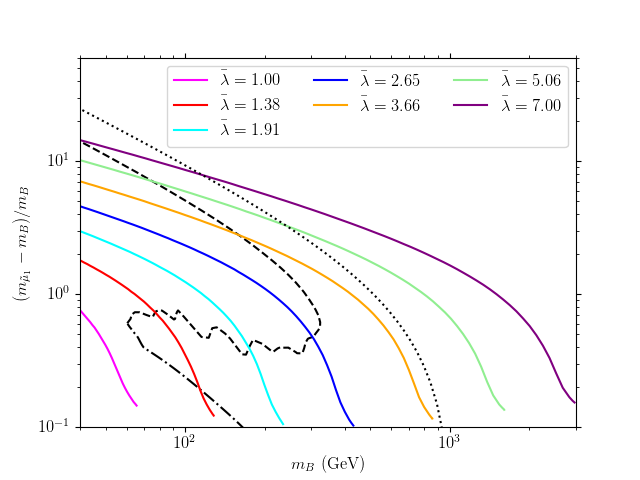} 
\caption{Parameter space for bino annihilation models which satisfy the DM relic density for fixed left-right smuon mixing angle $\theta_{\Tilde{\mu}} = - \pi /4 $ and smuon mass ratio $m_{\Tilde{\mu}_2}/m_{\Tilde{\mu}_1} = 1.25$. Each colored curve corresponds to a different multiple, $\bar{\lambda}$, of the bino-slepton-lepton coupling in the MSSM indicated in the legend and terminates at small relative mass splittings where bino annihilation contributes $\lesssim 80 \%$ of the effective annihilation rate. Collider constraints on the production of two mass degenerate chiral smuons with relatively large mass splittings from the bino~\cite{ATLAS:2019lff,ATLAS-CONF-2022-006} and a single chiral smuon for compressed spectra~\cite{LEPlimitsSlepton,ATLAS:2019lng} are show by the black dashed lines and dot-dashed, respectively. Assuming the left-right smuon mixing arises from a trilinear interaction with the SM Higgs boson, the dotted line shows the upper limit on the relative mass splitting for a given bino mass from EW vacuum stability corresponding to a smuon mass splitting parameter of $y \simeq 23$.} 
\label{fig:bulkRelic} 
\end{figure}

As discussed in ASU, the origin of the off-diagonal term in the smuon mass matrix which yields the left-right mixing need not be specified in order to have models with $\sim {\rm TeV}$ sleptons satisfying both the DM relic density and $g_\mu -2$. However, if we associate the left-right mixing with electroweak (EW) symmetry breaking in the SM, the corresponding trilinear interaction between chiral smuon eigenstates and the Higgs boson can have some impact on the phenomenology of the model. In particular, a large trilinear coupling can impact the calculation of the relic density in models where the effects of co-annihilation are important. In addition, large trilinear couplings in the scalar potential can destabilize the EW vacuum, yielding constraints from the lifetimes of the metastable EW vacua. In a detailed analysis of our model in ASU, we determined that in general a smuon mass splitting parameter of $y \gtrsim 23$ yields an EW vacuum metastable on timescales too short relative to the age of the Universe.

\begin{figure*}[t]
\includegraphics[width=0.49\linewidth]{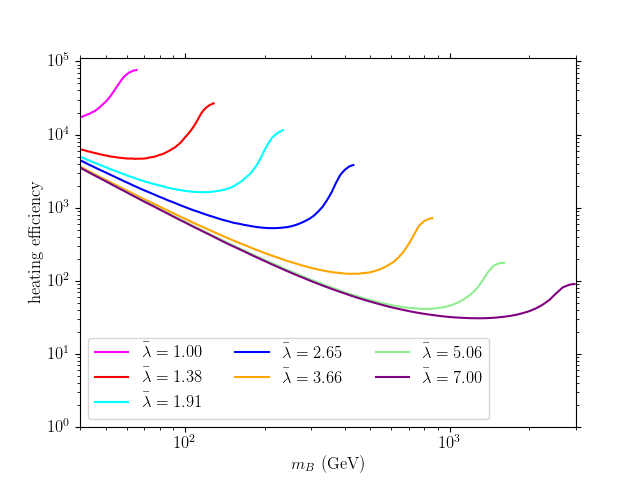}
\includegraphics[width=0.49\linewidth]{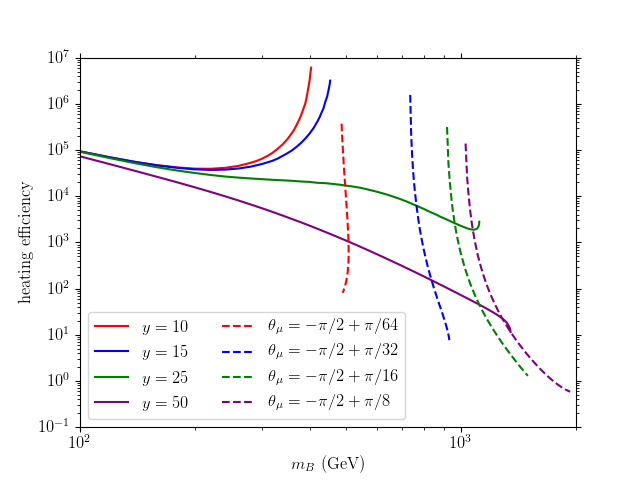}
\caption{Total NS heating efficiency for bino annihilation models shown in Fig.~\ref{fig:bulkRelic} (left, same color scheme) and smuon co-annihilation models along the RHB shown in Figs.~2 and~3 of ASU. Heating efficiency is normalized to the geometric capture rate, such that every DM particle incident on the NS would be captured after a single scatter for $f>1$. The NS equation of state is assumed to be BSk24~\cite{Pearson:2018tkr} with NS mass of $1.5 \, M_\odot$ and radius of $12.6 \, {\rm km}$.} 
\label{fig:totheating} 
\end{figure*}

We can separate the parameter space of the above DM model into two distinct regions characterized by the dominant processes which deplete the relic density. For the region of parameter space studied in ASU, the bino is nearly degenerate in mass with at least one of the sleptons such that co-annihilation processes control the DM relic density. In the ASU parameter space, we only consider coupling constants identical to the bino-slepton-lepton couplings in the MSSM (i.e. $\bar{\lambda}=1$) and assume the $CP$-violating phase vanishes. Accounting for $g_\mu -2$ requires the smuon mixing to be small for lighter $m_B$, yielding models along either a left- or right-handed branch (RHB) of the ASU parameter space. Satisfying the relic density with slepton co-annihilation then fixes the relative mass splitting between the lightest slepton and the bino along curves with either $y$ or $\theta_{\Tilde{\mu}}$ held constant.

For the parameter space in which the depletion of the relic density is controlled by bino annihilation, we consider both $\bar{\lambda} \gtrsim 1$ and $\phi_{\Tilde{\mu}} \neq 0$. We assume $\theta_{\Tilde{\mu}} = - \pi /4 $ since the $s$-wave contribution to the smuon-mediated bino annihilation to muons is maximized for larger left-right mixing. We also assume a fixed ratio of smuon masses $m_{\Tilde{\mu}_2}/m_{\Tilde{\mu}_1} = 1.25$. Since the bino annihilation cross section is largely independent of $\phi_{\Tilde{\mu}}$, we can then determine the smuon mass necessary to yield the correct relic density for a choice of bino mass and $\lambda_{\Tilde{\mu}_{R,L}} $. The $g_\mu -2$ contribution in this parameter space is typically too large to match the observed value without a near-maximal $CP$-violating phase, which suppresses the anomalous magnetic moment of the muon $\propto \cos \phi_{\Tilde{\mu}}$. For each model point in the bino annihilation parameter space we fix $\phi_{\Tilde{\mu}}$ to a value which yields the central observed value of $g_\mu -2$.

Using the DarkSUSY package~\cite{bringmann2018darksusy}, we determine the parameter space where bino annihilation can satisfy the observed DM relic density. We show the relative mass splittings between the lightest smuon and the bino as a function of bino mass for several benchmark curves with fixed values of $\lambda_{\Tilde{\mu}_{R,L}} $ in Fig.~\ref{fig:bulkRelic}. In order for the bino annihilation cross section to match that of thermal relic DM, heavier bino masses must be compensated by either a larger coupling or a smaller relative mass splitting between the lightest smuon and the bino. For the largest values of $\lambda_{\Tilde{\mu}_{R,L}} $, we see models where the relic density can be depleted to the observed value through bino annihilation mediated by $\sim {\rm TeV}$ smuons. For smaller relative mass splittings, $\lesssim 0.2$, we see the effects of co-annihilation start to allow for models with even heavier bino masses to satisfy the relic density.

Although a detailed interpretation of collider constraints on slepton masses~\cite{Dutta:2014jda,Han:2014aea,Dutta:2017nqv} for our particular model is beyond the scope of this work, we also show the constraints from LEP and LHC in Fig.~\ref{fig:bulkRelic} assuming the production of either a single right~\cite{LEPlimitsSlepton} or left~\cite{ATLAS:2019lng} chiral smuon or two mass degenerate chiral smuons~\cite{ATLAS:2019lff,ATLAS-CONF-2022-006}, each decaying to a bino and a muon. Even though these constraints should only be considered indicative, we can see that all of the model curves except for the MSSM benchmark coupling can yield viable parameter points satisfying the DM relic density. Also, note that indirect detection searches can in principal constrain DM models with $m_B \lesssim 100 \, {\rm GeV}$ and a thermal WIMP annihilation cross-section, however WIMP annihilation to muons can be consistent with multimessenger constraints for $ m_B \gtrsim 40 \, {\rm GeV}$~\cite{DiMauro:2021qcf,Abdughani:2021oit}. Regarding slepton co-annihilation models, neither current collider searches nor indirect detection experiments are sensitive to parameter space considered in ASU. However, a future lepton collider with a $\gtrsim 500 \, {\rm GeV}$ center of mass energy~\cite{deBlas:2018mhx,Berggren:2013vna,Baum:2020gjj} would be a complementary probe, potentially able to distinguish between the slepton mediator models explored in this Letter and other DM models which can be observed by the kinetic heating of NSs.  

Conventional direct detection experiments designed to detect weak scale dark matter scattering off nuclei have limited sensitivity to Majorana singlet DM candidates which only couple to the SM at tree-level through muons~\cite{Sandick:2016zut,Acuna:2021rbg,Hamaguchi:2022wpz}. More specifically, the leading contribution to nucleon scattering typically arises from the anapole moment. In addition to being loop-suppressed, the anapole moment only yields proton scattering rates large enough to be detected at next generation direct detection experiments in scenarios with very small mass splittings between the lightest smuon and the bino. 

We have checked that none of the models shown in Fig.~\ref{fig:bulkRelic} would be accessible to an experiment with sensitivity similar to what has been proposed for DARWIN~\cite{aalbers2016darwin}.
While the parameter space of the co-annihilation models considered in ASU can feature sufficiently compressed spectra to enhance the anapole moment, a future experiment similar to DARWIN would only be sensitive to a small subset models with MSSM-like $\lambda_{\Tilde{\mu}_{R,L}} $ and $m_{B} \simeq m_{\Tilde{\mu}_1} \lesssim 400 \, {\rm GeV}$~\cite{Acuna:2021rbg}. 

\medskip
{\it Neutron star kinetic heating.}---As an alternative to conventional direct detection experiments, the effects of DM scattering can also be detected in observations of old NSs. More specifically, the transfer of kinetic energy from the incident DM flux to the neutron, proton, muon and electron NS constituents can result in an elevated NS temperature which could be measured by JWST. Following Refs.~\cite{Joglekar:2020liw,Bell:2020jou,Bell:2020lmm}, we assume the BSk24 family of configurations for the NS equations of state (EoS)~\cite{Pearson:2018tkr}. We calculate the associated DM capture rate for all potentially significant NS constituents given the model with bino-slepton-lepton interactions described by Eq.~\ref{eq:Lint} and the subsequent discussion of the parameter space in which both the DM relic density and $g_\mu -2$ are satisfied. 

We calculate the DM capture rate summed over the scattering contributions from all NS constituents (see Supplementary Material for discussion of contributions from scattering off different targets) for the models along the curves in Fig.~\ref{fig:bulkRelic} and on the RHB of the models satisfying the relic density and $g_\mu -2$ in Figs.~2 and~3 of ASU. We characterize the kinetic heating of NSs by the heating efficiency, $f$, which is the ratio of the capture rate to the `geometric capture rate' for a $1.5 \, M_\odot$ NS with a radius of $12.6 \, {\rm km}$ (BSk24-2 EoS from Refs.~\cite{Bell:2020jou,Bell:2020lmm}). For $f>1$ all DM particles incident on the NS will be captured after a single scatter, while $f<1$ gives the probability for incident DM particles to be captured~\cite{Joglekar:2020liw}. Models with $f > 1$ would heat NSs to $\sim 1600 \, {\rm K}$ over $\sim 10^9$ years given a dark matter density equivalent to that at the solar radius, $\sim 0.4 \, {\rm GeV}/{\rm cm^3}$. NS surface temperatures $\gtrsim 1000 \, {\rm K}$ are potentially detectable by JWST and significantly higher than the $\sim 100 \, {\rm K}$ expected from the cooling of a old NS~\cite{Yakovlev:2004iq,Page:2004fy}. The smallest NS temperature conceivably detectable in future infrared observations of ${\cal O}(100)$ old NSs corresponds to $f\sim 10^{-2}$~\cite{Baryakhtar:2017dbj,Raj:2017wrv}.

In the left panel of Fig.~\ref{fig:totheating}, we see that the lower slepton and bino masses associated with smaller $\lambda_{\Tilde{\mu}_{R,L}} $ in bino annihilation models can yield $f \gg 1$ even in the parameter space where the relative mass splitting between the bino and lightest smuon is large. For increasingly larger $\lambda_{\Tilde{\mu}_{R,L}} $ towards the perturbative unitarity limit, the heating efficiency decreases as $m_{B}$ increases past the $\sim {\rm TeV}$ scale but $f \gtrsim 1$ for all models investigated. We also see that the heating efficiency for all of the benchmark $\lambda_{\Tilde{\mu}_{R,L}} $ curves becomes strongly enhanced as the bino and lightest smuon become mass degenerate. Although $\phi_{\Tilde{\mu}} \sim \pi /2$ for these models to satisfy $g_\mu -2$, the dominant contributions to NS heating are virtually independent of the $CP$-violating phase.

In the right panel of Fig.~\ref{fig:totheating}, the smuon co-annihilation model benchmarks with fixed $y$ demonstrate the transition from cases more similar to bino annihilation, in which the relative mass splitting must decrease to compensate for increasing $m_{B}$, to cases with larger $y$, in which the relative mass splitting must remain constant or increase to compensate for the associated enhancements to the effective annihilation rate. In particular, we see the NS heating efficiency fall with $m_{B}$ in the cases with larger $y$. These effects in the smuon co-annihilation model parameter space are most clearly demonstrated for the benchmark curves with fixed $\theta_{\Tilde{\mu}}$, for which we see the heating efficiency falling rapidly as $y$ increases along the curves towards larger $m_B$. However, we see that all models along the RHB should heat NSs to temperatures sufficient to be detected by future infrared observations.

\begin{figure}[t]
\includegraphics[width=\linewidth]{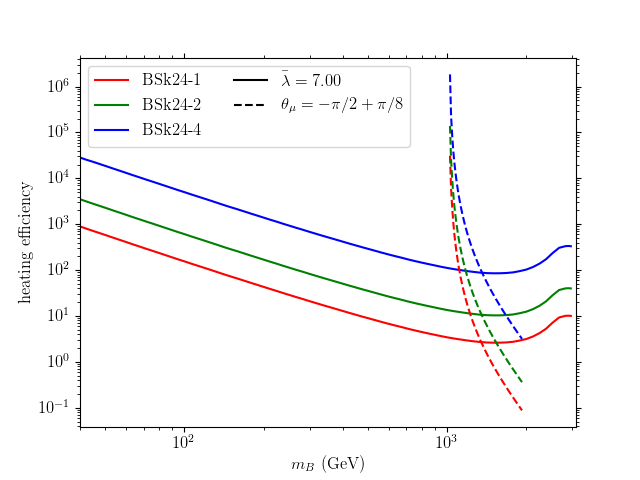}
\caption{Total NS capture efficiency for bino annihilation models with $\bar{\lambda} = 7$ from Fig.~\ref{fig:bulkRelic} (solid) and smuon co-annihilation models with fixed $\theta_{\Tilde{\mu}} = - \pi /2 + \pi/8 $ from ASU (dashed) given different EoS configurations BSk24-1 (red), BSk24-2 (green) and BSk24-4 (blue), as described in the text.} 
\label{fig:EOS} 
\end{figure}

In Fig.~\ref{fig:EOS}, we show the total capture efficiency for bino annihilation models with the largest $\bar{\lambda} = 7$ in Fig.~\ref{fig:bulkRelic} and smuon co-annihilation models with fixed $\theta_{\Tilde{\mu}} = - \pi /2 + \pi/8 $ given different NS EoS configurations. In addition to BSk24-2 which we assume as our benchmark EoS configuration, BSk24-1 and BSk24-4 from Refs.~\cite{Bell:2020jou,Bell:2020lmm} respectively correspond to a mass of $1.0 \, M_\odot$ with a radius of $12.22 \, {\rm km}$, and a mass of $2.16 \, M_\odot$ with a radius of $11.97 \, {\rm km}$. We can see that, even under the assumption of the BSk24-1 configuration yielding the least efficient DM capture, the model curves with the smallest capture efficiencies in Fig.~\ref{fig:totheating} would heat NSs to sufficiently high temperatures to potentially be detected.

\medskip
{\it Conclusions.}---In this Letter, we have studied dark matter models which can satisfy $g_\mu -2$ with interactions mediated by scalar lepton partners. We investigate a representative parameter space of models with Majorana singlet dark matter produced thermally in the early Universe either by dark matter annihilation or co-annihilation processes involving the scalar leptons. For such models with dark matter interactions mediated by scalar muon partners, the only tree-level interactions are with muons and are notoriously difficult to probe at either indirect detection or collider searches. In contrast to similarly challenging searches at conventional direct detection experiments, the presence of leptonic constituents makes the kinetic heating of neutron stars particularly appealing. 

For nearly the entire parameter space investigated, we show that scattering is efficient enough for the entire dark matter flux incident on the neutron star to be captured. Geometric capture efficiency of dark matter in the local galactic halo implies the kinetic heating of a population of old neutron stars to temperatures which should observable by future infrared telescopes, such as JWST. The observation of an old population of neutron stars with temperatures $\gtrsim 1000 \, {\rm K}$ would thus be consistent with scalar muon mediated interactions, but without a way to distinguish from other dark matter models which also have sufficiently strong interactions with neutron star constituents. On the other hand, non-observation of neutron stars which have undergone kinetic heating would definitively rule out Majorana singlet dark matter models satisfying $g_\mu -2$ with scalar muon partners up to the $\sim {\rm TeV}$ scale.

\begin{acknowledgments}
JTA would like to thank IFPU, Trieste and the National Institute of Physics, UP Diliman, where most of this work was carried out, for the hospitality and support. This work was supported by the research grant ``The Dark Universe: A Synergic Multi-messenger Approach" number 2017X7X85K under the program PRIN 2017 funded by the The Italian Ministry of Education, University and Research (MIUR), and by the European Union's Horizon 2020 research and innovation program under the Marie Sklodowska-Curie grant agreement No 860881-HIDDeN. Partial funding is also provided by the National Science and Technology Council of the Republic of China (formerly the Ministry of Science and Technology), with grant number MOST 111-2811-M-007-018-MY2.
\end{acknowledgments}

\bibliographystyle{kp.bst}
\bibliography{bulkNS-final-arxiv}

\clearpage

\begin{center}
 {\LARGE SUPPLEMENTARY MATERIAL}
\end{center}

\begin{figure}[h]
\includegraphics[width=\linewidth]{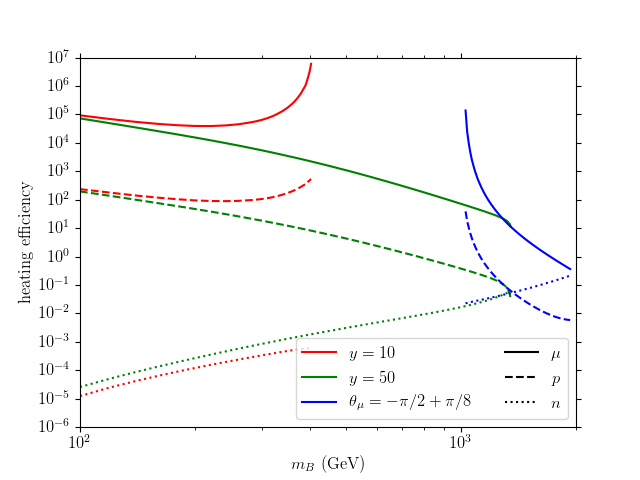}
\caption{Capture efficiency contributions from muon (solid), proton (dashed) and neutron (dotted) scattering for model curves with fixed $y=10$ (red), $y=50$ (blue) and $\theta_{\Tilde{\mu}} = - \pi /2 + \pi/8 $ (green) from the right panel of Fig.~\ref{fig:totheating}.} 
\label{fig:targetheating} 
\end{figure}

In contrast to the anapole moment, loop-induced couplings of the bino to nucleons through both Higgs and Z-boson exchange are less sensitive to the mass splitting between the lightest smuon and the bino, but only the former can become important at conventional direct detection experiments in models with a sufficiently large trilinear coupling between the Higgs and the smuons (see discussion in ASU). For the kinetic heating of NSs in our slepton mediator model, we demonstrate that the contribution from nucleon scattering at loop-level is always subdominant to that of tree-level muon scattering. Note that all nucleon scattering contributions are calculated in an effective contact operator framework with Wilson coefficients defined in Ref.~\cite{solonDD} and nucleon form factors from Ref.~\cite{Bishara:2017pfq}.

In Fig.~\ref{fig:targetheating}, we show the contributions to NS heating from DM scattering off different NS constituents for several model curves from the right panel of Fig.~\ref{fig:totheating}. For $y=10$, we see that the NS heating contribution from muon scattering dominates over the nucleon scattering contributions by several orders of magnitude despite the number density of the muons within the NS being less than that of the protons and much less than that of the neutrons. In fact, with the proton scattering contributions also much larger than the neutron scattering contributions due to the enhanced anapole moment, we see there the hierarchy of NS heating contributions is actually the reverse of that of the number densities of NS constituents~\cite{Bell:2019pyc,Garani:2019fpa,Joglekar:2019vzy,Joglekar:2020liw}. A similar ordering of NS heating contributions can be seen across most of the parameter space of bino annihilation models, although the ratios of respective contributions are not as large as in the smuon co-annihilation models due to relatively uncompressed spectra.

While the reverse hierarchy of NS heating contributions holds through most of the parameter space for both bino annihilation and smuon co-annihilation models, we can see the nucleon scattering become more significant as the smuon mass splitting parameter becomes large. In particular, the hierarchy of NS heating contributions begin to change for smuon co-annihilation models with fixed $y=50$, as the neutron scattering contribution becomes similar to the proton scattering contribution at $m_B \gtrsim 1 \, {\rm TeV}$. The mass splitting between the bino and lightest smuons can remain relatively large for these points, thus suppressing the anapole moment and the muon scattering contributions. When $y > 50$ along the model curve with fixed $\theta_{\Tilde{\mu}} = - \pi /2 + \pi/8 $, we can see the neutron scattering contribution becomes larger than the proton scattering contribution and within an order of magnitude of the muon scattering contribution.

\end{document}